\documentclass{article}

\usepackage[english]{babel}

\usepackage[letterpaper,top=2cm,bottom=2cm,left=3cm,right=3cm,marginparwidth=1.75cm]{geometry}

\usepackage{amsmath}
\usepackage{graphicx}
\usepackage[colorlinks=true, allcolors=blue]{hyperref}

\title{High stakes classification with multiple unknown classes based on imperfect data}
\author{Haakon Bakka}

\begin{document}
\maketitle

\begin{abstract}
High stakes classification refers classification problems where erroneously predicting the wrong class is very bad, but assigning ``unknown'' is acceptable. We make the argument that these problems require us to give multiple unknown classes, to get the most information out of our analysis. With imperfect data we refer to covariates with a large number of missing values, large noise variance, and some errors in the data. The combination of high stakes classification and imperfect data is very common in practice, but it is very difficult to work on using current methods.

We present a one-class classifier (OCC) to solve this problem, and we call it NBP. The classifier is based on Naive Bayes, simple to implement, and interpretable. We show that NBP gives both good predictive performance, and works for high stakes classification based on imperfect data.

The model we present is quite simple; it is just an OCC based on density estimation. However, we have always felt a big gap between the applied classification problems we have worked on and the theory and models we use for classification, and this model closes that gap. Our main contribution is the motivation for why this model is a good approach, and we hope that this paper will inspire further development down this path.

%TODO LIST:
%Spell check. 
%Search todos.
%Note: Should ``prediction weights" be replaced with another word+symbol everywhere?
%I could use likelihoods $l_c$ and $\mathcal L_c$ $ L_c$ $\mathcal l_c$..?

\end{abstract}

\section{Introduction}

\subsection{Application example} \label{sec:introapp}

When farmed salmon escapes from captivity and mixes with wild salmon, this is a big problem for the genetic diversity of the wild salmon. 
In order to reduce the number of escapes, the industry captures escaped salmon and wants to track it back to the location from which it escaped.
This can be done with a mix of genetic analysis and by looking at the element compositions in the layers of the fish scales.
We will only consider the element composition approach.

Fish scales are built up layer by layer over the months that the fish is alive.
Samples from these layers can be analysed with mass spectrometry to find the composition of rare elements, like Strontium and Barium.
This composition can be used to identify where the fish was located while the layer was formed.
Reference data are collected from all locations (salmon farms) under suspicion, in order to have something to 
compare the escaped fish to.

Unfortunately, the element composition data are very noisy, have many missing values, and contain some errors.
This is partly because extracting the samples and analysing them is a complicated process.
And, partly because the biochemical processes that relates the element composition of the environment to the element composition of the fish scale layer are complex and varies in time.

The dataset consists of a set of observations of element concentrations $x_{i,j}$ from scale $i$ and element $j$, and a location $y_i$ where the fish was sampled from.
One sample from an escaped salmon is a set of element concentrations $z_j$.
The goal is to predict which location this salmon escaped from.

When the prediction is used by decision makers to blame one location for letting the salmon escape, it is much better to answer ``we do not know" than to give an incorrect prediction. 
In this sense this is a high stakes classification problem.
However, if we can exclude some locations, this is much more useful for further investigations than concluding ``we do not know anything".

If there are two possible candidate locations, for which the element composition of the escaped salmon can plausibly come from either location, we do not want to exclude any of them.
Hence, if one location is much more ``likely" than another, but both are ``clearly possible", we do not want to pick the more likely location, but to predict that it can be either of the two locations.

In the rest of the introduction, we will frame this problem in terms of statistics and machine learning.

\subsection{General problem framework}

We have training data $(X, Y)$,
where $X$ is a matrix of discrete and continuous observations, $X=(\vec x_i)_i$, referred to as covariates, and $Y=(y_i)_i$ is the class of each observation row.
We use $X = (X_j)_j$ to refer to the different covariates, where $X_1$ is the first covariate column, with all the observations of this covariate.
The $y_i$ can take values $1, 2, ..., C$ representing $C$ different classes.
Probabilistic models for the joint distribution $(X, Y)$ are usually built from conditional distributions $Y|X$, e.g. regression, or $X|Y$, e.g. Naive Bayes.

We have prediction data consisting of a single vector $\vec z$ where the corresponding $y^*$ is unknown.
The goal is to predict the value of $y^*$, and the quality of an algorithm can be analysed e.g. by standard cross-validation, using e.g. accuracy of point predictions, a proper scoring rule, and/or some analogue to coverage of confidence intervals.
The application context determines the 
most appropriate measures of prediction quality.
For simplicity, we consider just one prediction. 
When there are several predictions to be made, we assume they are predicted one at a time, independently from each other.

In this paper we mainly consider situations where there might not be enough information in the data to reliably predict the class. We give a negative example: this paper is not motivated by checking whether an image of an animal is of a dog or not, because you would expect the information to exist in the image. 
We are mainly concerned with measurements of any system where we do not know whether there is enough information available to make reliable predictions.

Prediction with uncertainty usually takes the form of probabilities, likelihoods, or weights $w_c$ that $y^*$ belongs to class $k$.
Point predictions are often defined as a selection of one class, but we make a different consideration as follows.
We consider a point prediction to be a binary vector $W=(W_1, ..., W_C)$ predicting whether $y^*$ can belong to the different classes.
E.g. $W=(0, 1, 1, 0)$ means that $y^*$ can belong to class 2 and 3 but not to class 1 and 4.
If the distributions $X_2|Y$ and $X_3|Y$ are too similar, this would be a good result.
For high stakes classification, we need to provide point predictions in this way, and not give whichever class happens to fit slightly better.
Another alternative would be to have only one unknown class in the model, and reduce e.g. $W=(0, 1, 1, 0)$ to $W=\text{Unknown}$, but this would throw away information that can be very useful in decision making.

\subsection{Concluding vs excluding}

An attribute of these applied problems is that you can often exclude the possibility that a prediction observation is from a given class, but that it is more difficult to conclude that it must be from one specific class.
Consider for example a location where the element composition always contain very small amounts of Barium. 
An escaped salmon with large amounts of Barium in its scales cannot possibly come from that location.
If one location has small amounts of Barium, similar to the escaped salmon, it might be that the salmon escaped from there, but it can always have been from some other location instead.
Even if all the covariates fit well for a location, it is technically possible that another unknown location exists with the same profile, or another profile that matches the escaped salmon. 

At some point a decision maker has to decide based on expert knowledge that another unseen candidate is unlikely.
If we assume that all classes are included in the dataset, and that we have a rich set of covariate samples from all the classes, and W has exactly one 1, we can conclude that the prediction belongs to that class.

In Section \ref{sec:introapp} we describe a situation where, for the prediction, we would heavily penalise an approach exluding a location that is actually possible.
Hence, we want the cutoff when producing $W$ from $w$ to be small.
However, there may be applications, or a context of this applied problem, where we would put a heavier penalty on not excluding the class.
For example, if every location that is not excluded suffers some economic penalty.

\subsection{Modelling challenge}

In addition to the problem framework described so far, we have the following main methodological challenge.
\emph{Challenge 1: If the $z$ fits better with one distribution than another, but is still plausible under both distributions, we want $W_c$ to be 1 for both the corresponding classes.}
Many models do not satisfy Challenge 1.
We show the issue with Naive Bayes in Section \ref{sec:NBissue}.
%However, linear regression with a strict cutoff on the predicted probability can in some cases satisfy Challenge 1, as it only models the location parameter, not the spread. However, the cutoff for W might have to depend on the data.

Let us make a hypothetical scenario, with two independent covariates $X_1$ and $X_2$, and two classes $A$ and $B$.
Assume class A has a unimodel distribution for each covariate, decaying away from the mode.
Assume the covariates of class B has the same distributions as for class A, but that the covariates are more concentrated around the mode, i.e. their marginal standard deviations are smaller.
Then if the prediction has covariates fitting within class B, they also fit with class A. 
Most models would predict that class B is the correct class,
because a tighter spread gives a higher likelihood. 
However, 
we do not want this to result in a conclusive statement that the prediction belongs to class B.

\subsection{One-class classifiers}
A natural approach to solve this modelling challenge is a one-class classifier (OCC).
An OCC produces for a class $C$ a 0/1 prediction whether $\vec z$ belongs to the class or not.
By fitting the OCC separately to each class, we get the desired point prediction.
E.g. for $W=(0, 1, 1, 0)$ we have fitted four OCCs.

The added assumption with the OCC is that we do not use what we infer about the distribution of $X|c$ for one class when we infer the distribution for another class. 
Making this assumption is desirable in order to resolve Challenge 1, 
because then the increase of $w$ of one class cannot decrease the $w$ of another class.
The new model we present is an OCC.
We do not want an OCC with a hard cutoff, e.g. with hyperspheres around location parameters. This would only give us $W$, but we also want $w$.

%\subsection{Literature context}

%TODO: Find references and discuss them.

\section{Statistical motivation and background}

\subsection{Naive Bayes}

In Naive Bayes (NB), we model the joint distribution $\pi(X, Y) = \pi(X|Y) \pi(Y)$.
In order to simplify notation, we assume that the prior density for the class memberships $\pi(Y)$ is uniform.
We make this assumption for the rest of the paper, as it is simple to generalise back to non-uniform $\pi(Y)$.

The first step is to fit the distribution $X|Y$.
In NB we fit each covariate $X_j$ with a separate distribution, for each class, to get the density function $f_{X_j|Y=c}$.
%, and the distribution (cumulative density) function %$F_{X_j|Y=c}$.
Then
$$\pi(X|Y=c) = \prod_{j=1}^J f_{X_j|Y=c}.$$

For predicting $y^*$ from $z$, 
the likelihood (prediction weights $w$) are 
$$\pi(Y=y^* | X=z) = \frac{\pi(X=z|Y=y^* ) \pi(Y=y^*)}{\pi(X=z)} \propto \pi(X=z|Y=y^* ).$$
Hence the prediction weights are 
$$w_c = \prod_{j=1}^J f_{X_j|Y=c}(z_j).$$
However, because of the proportionality, the weights themselves have no inherent meaning, only their relative size does.
E.g. $w=(1, 0.1, 5, 7)$ is equivalent to $w=(3, 0.3, 15, 21)$.

To produce our binary $W$, we can threshold the relative differences between weights. E.g. we can say that any weight less than 1/10th of the largest weight means that the class is impossible.
For $w=(1, 0.1, 5, 7)$, this would mean that $W = (1, 0, 1, 1).$
This threshold can be set by cross validation.

\subsection{The issue with NB} \label{sec:NBissue}

\begin{figure}
\centering
\includegraphics[width=0.7\textwidth]{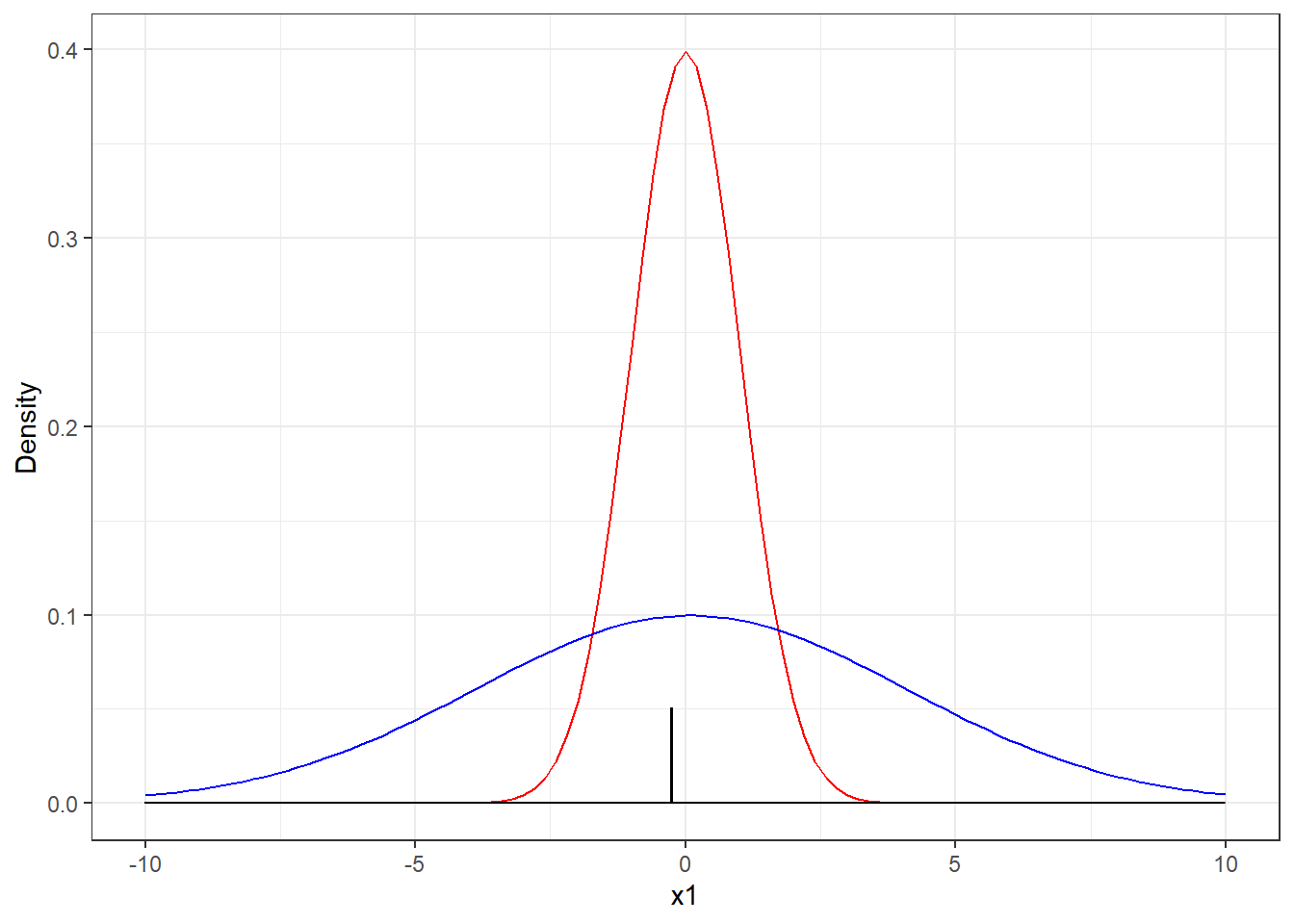}
\caption{\label{fig:illustr1}This
figure illustrates two $X|Y$ densities in a Naive Bayes model, for two different classes.
The black line represents a new observation at $z=-0.25$, for which we want to predict class membership.
The density according to the red density (the red class) is 0.387, while the density according to the blue class is 0.0994.
The ratio between the two densities is 3.9, which means that Naive Bayes would predict that the new observation belongs to the red class, with ca 80\% 
probability.}
\end{figure}

NB is flawed in the case of our application example and it does not solve Challenge 1.
We illustrate this in Figure \ref{fig:illustr1}.
Here, we see a case where the observation $z$ is a likely outcome under both distributions.
But, since it fits much better with the narrow distribution, NB will strongly conclude that it must come from this distribution, and hence the corresponding class.
In higher dimensions this flaw is substantially worse.

%NB is quite good with missing data.
%For the fit, the missing data can be taken out of each covariate separately.
%For the prediction, the missing value can be taken out of the product in Equation (first eq in 2.1).
% TODO: Is this true, can it just be taken out?? Probably not! If all the densities of that class is way less than 1, taking it out changes everything a lot!!!! Ooops 
%NB struggles with outliers.
%An outlier results in a very large relative density, and results in an incorrect prediction that is very confident.

\section{NBP: A variant of Naive Bayes with p-values}

The previous sections motivate why we want to create a new OCC model.
We build this model based on the NB model, in the sense that we fit the density function
$f_{X_j|Y=c}$, from the training data, assuming some distribution family (e.g. Gaussian).
From the density function we compute the cumulative density function $F_{X_j|Y=c}$, also known as the distribution function.

The change from the NB model is that we do not use the density, but the distribution. We use the distribution to compute exceedance probabilities, i.e. p-values.
The p-value $p_{j, c}$ belonging to covariate $j$ and class $c$ is the probability that $X_j|Y=c$ takes a value further out in the tail of the distribution than the prediction covariate $z_j$.
We use the two-sided p-value, hence, for $z_j$ smaller than the median,
$$p_{j, c} = 2*F_{X_j|Y=c}(z_j),$$
and similarly, with $1-F$, when $z_j$ bigger than the median.
This p-value is always in $[0, 1]$, and a larger p-value signifies an increase in the likelihood that $Y$ is the class $c$.

Since we never use the density $f$, only the distribution $F$, it is natural to suggest estimating $F$ directly. 
We did not try this yet.
Hence, to compute small p-values we heavily rely on the parametric assumptions (the chosen family of distributions), and the choice of estimators of these parameters.

\subsection{Solving Challenge 1}

Consider the distributions in Figure \ref{fig:illustr1}.
The p-values for $z$ in this case is
0.80 for the red class, and 0.93 for the blue class.
From these p-values the only reasonable conclusion is that $z$ could have come from both models.
In this example $w = (0.80, 0.93)$, and $W=(1,1)$ for any reasonable cutoff.

The p-values do not indicate which of these two classes ``fit better". This is different from how it was with the densities in the NB model.
The one that ``fit better" according to NB is the one with a smaller p-value.
There is no inherent comparison between the two fits, since the NBP is an OCC.
However, when the p-values become small, e.g. $<0.05$, the better fitting distribution according to NB usually has a smaller p-value.
%Hence we can not use the p-values to compare which class was the most likely to have .....?
% What if P values are very low, then they work!

\subsection{Combining p-values}

To compute the prediction weights $w$, we need to aggregate the p-values over $j = 1,...,J$. 
If we think in terms of exceedance probabilities and cumulative distributions, it is not clear what to do, but if we think in terms of p-values it is much easier. 
This is the reason we call the new model NBP.

A p-value relies on the concept of ``stranger than z", the set of covariate-values that are ``more out". 
In the 1-dimensional case this is straight forward; any value less than ``z below median" or greater than ``z above median" are ``more out".
For several dimensions it is less clear.
E.g. for two dimensions, we can think about ``outside the rectangle" or ``outside the ellipse".

In this paper we choose Fisher's method for combining p-values.
Let
$$ G_c = -2 \sum_{j=1}^J \log(p_{j, c} ),$$
where our $p_{j, c}$ are independent across $j$ (for a fixed $c$).
If we assume all the underlying null hypotheses are true, then $G_c$ has a chi-squared distribution with $2J$ degrees of freedom ($\mathcal X^2_{2J}$).
From this we can compute the total p-value, denoted $w_c$, as one minus the distribution function for $G_c$.
These $w_c$ are the prediction weights we want to compute.

How to go from $w$ to $W$, which cutoff to choose when making the predition binary, will always depend on the application context. 
Two different high stakes classification have different tradeoffs of how many ``unknown" predictions are worth having one extra erroneous conclusion.

One possibility would be to set up $W^{(1)}$ to $W^{(4)}$, with four different cutoffs, e.g. use 1.98E-09, 6.33E-5, 0.005, 0.05. 
These values correspond to different common values used to think about significance and p-values.
The two smallest numbers here correspond to outside of 6 standard deviations and outside of 4 standard deviations on a Gaussian distribution.
In our experience, when presenting the results to stakeholders and decision makers, it quickly becomes clear if one is using a very inappropriate cutoff.

Another possibility would be to define the utility of predictions, e.g. by defining that 20 unknown-predictions are equivalent to 1 incorrect prediction, and then perform cross-validation to estimate the cutoff.

\subsection{Limiting effect of outliers}

In larger datasets there are often some recorded numbers that are incorrect. If they are off by orders of mangitude, one can remove them by cleaning the data. But if they are somewhat reasonable, e.g. off by a factor of two, we might not be able to detect them.
In this case, the best we can do is avoiding that single covariate values completely determine the prediction.

When fitting the distributions, we can use quantiles to estimate parameters, or we can remove extra outliers before estimating.
E.g. for the Gaussian distribution we can use the median to estimate the Gaussian mean parameter, and we can use quantiles to estimate the standard deviation parameter.

For the predictions, we have several $X_j$, and if one of them is incorrectly recorded we can avoid that this one completely determines the prediction $W$ in the following way.
Before we combine p-values, we can threshold p-values to not be larger than e.g. 6.33E-5.
For a Gaussian distribution this is equivalent to saying that ``if the $z_j$ are 4 standard deviations or more out from the median, set them to be exactly 4 standard deviations out".
Then, we can make the cutoff for constructing W slightly smaller than this threshold.

\section{Code example for the iris dataset}

\begin{figure}
\centering
\includegraphics[width=0.7\textwidth]{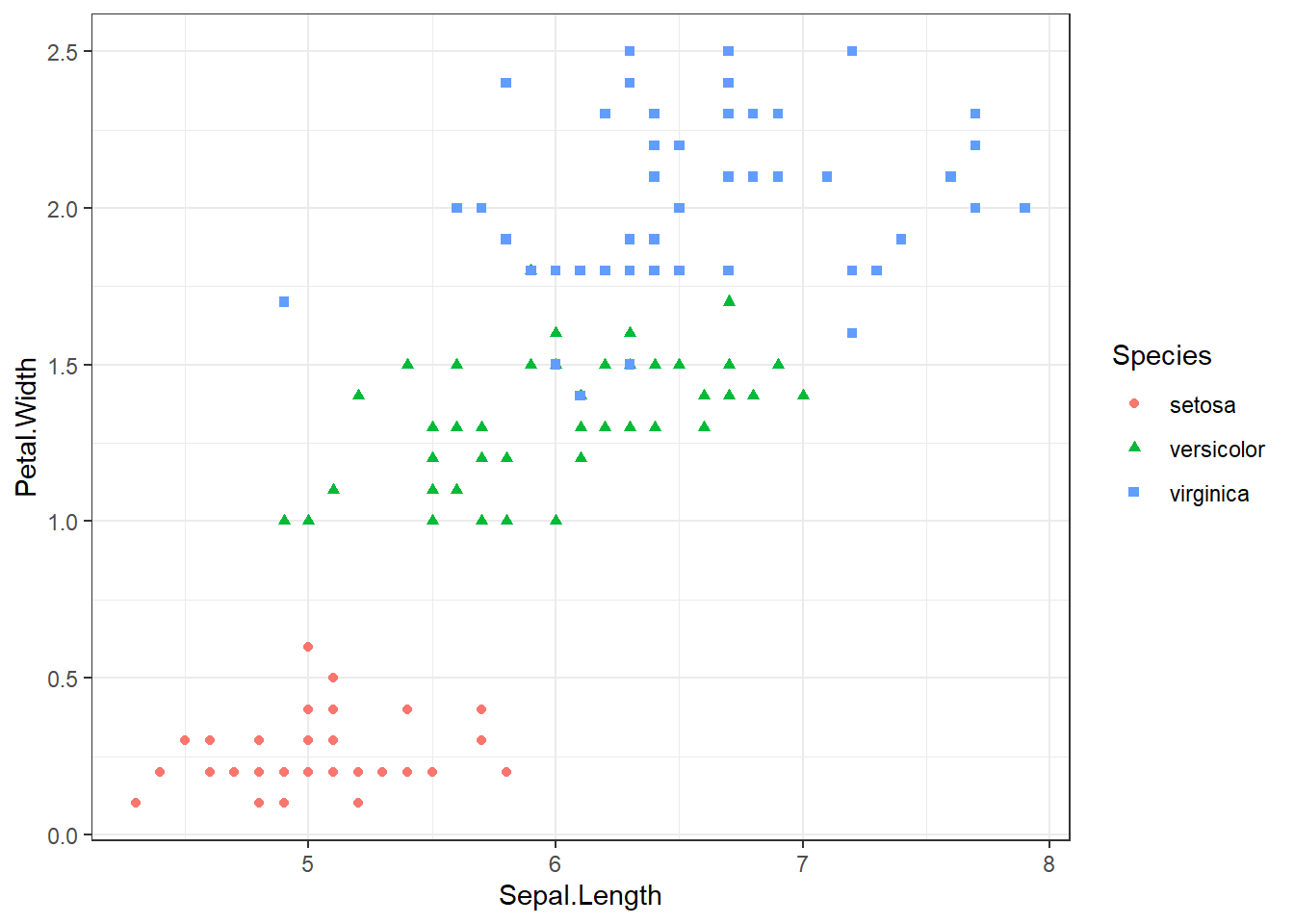}
\caption{\label{fig:btopic136-pwsl}This
figure shows two of the covariates of the iris dataset, together with class membership (color/shape).
In \emph{experiment 1}, the single red point with a value of 0.6 for Petal.Width is changed to have a Petal.Width value of 1.5.}
\end{figure}

We use the standard iris dataset as a brief example.
This dataset has 3 classes (setosa, versicolor, and virginica), and four covariates (Sepal.Length, Sepal.Width, Petal.Length, and Petal.Width).
Figure \ref{fig:btopic136-pwsl} shows a scatterplot of two of the covariates, with the point color showing the class.
We focus on the point with a Petal.Width value of 0.6. 
When holding out this point and predicting it, both NB and NBP are able to predict it to the correct class (setosa).
This is not surprosing, as the figure shows a very clean separation between the classes with only two of the four covariates.

\emph{Experiment 1.} We replace the Petal.Width value of 0.6 with 1.5. 
This illustrates how data can contain errors which are hard to detect; the value of 1.5 is still a very common value in the dataset and would not raise any red flags.
Then, NB predicts the wrong class versicolor, and NBP the correct class (setosa).
What happens is that three of the covariates suggest the correct class, but the erroneous covariate very strongly suggests that it cannot belong to the correct class.
The total prediction from NB gets steered too much from the single erroneous covariate.
NBP uses the suggested cutoff of 6.33E-5,
which means that the erroneous covariate gets a large impact, but not large enough to silence the three good covariates.
For more details, see supplementary materials.

\section{Element composition of escaped salmon}

We have provided the dataset in the R-package NBPclassify.
The supplementary materials contain more information, and the code for how to 
analyse this dataset with the NBP model.

\subsection{Prediction quality of NBP}

\begin{table}
\centering
    \begin{tabular}{r|r|r|r|r}
    	& A & D & M & T   \\ \hline
    	A & 100 & 0 & 0 & 0   \\ \hline
    	D & 0 & 89 & 11 & 0   \\ \hline
    	M & 0 & 21 & 79 & 0   \\ \hline
    	T & 4 & 0 & 6 & 90   \\ 
    \end{tabular}
    \caption{Confusion matrix for the classical NB. Each row represents 100 simulated 5-points hold-out, and the column name shows the predicted 
    class.}
    \label{tab:NB1}
\end{table}

\begin{table}
\centering
    \begin{tabular}{r|r|r|r|r}
    	& A & D & M & T   \\ \hline
    	A & 100 & 0 & 0 & 0   \\ \hline
    	D & 0 & 92 & 7 & 1   \\ \hline
    	M & 0 & 10 & 86 & 4   \\ \hline
    	T & 2 & 0 & 1 & 97  \\ 
    \end{tabular}
    \caption{Confusion matrix for the robust NB. Each row represents 100 simulated 5-points hold-out, and the column name shows the predicted 
    class.}
    \label{tab:NB2}
\end{table}

\begin{table}
\centering
    \begin{tabular}{r|r|r|r|r}
    	& A & D & M & T   \\ \hline
    	A & 100 & 0 & 0 & 0   \\ \hline
    	D & 0 & 89 & 3 & 8   \\ \hline
    	M & 0 & 7 & 46 & 47   \\ \hline
    	T & 0 & 0 & 0 & 100   \\ 
    \end{tabular}
    \caption{Confusion matrix for the NBP. Each row represents 100 simulated 5-points hold-out, and the column name shows the predicted 
    class.}
    \label{tab:NBP}
\end{table}

In order to consider NBP as a useful alternative to NB we have to show that the point prediction ``choose one class" is reasonably good.
For NBP we pick the class that has the highest p-value, and
for NB we pick the class with the highest density.
We also improved NB for this dataset, by defining robust estimators for the parameters,
and we refer to this improvement as robust NB.
The confusion matrices are in tables \ref{tab:NB1}, \ref{tab:NB2}, and \ref{tab:NBP}.
By comparing these tables, we see that the 
three algorithms all perform reasonably well.
The NBP is better at some classes, and worse at others, compared to the two NB algorithms.
Hence, NBP is a reasonable alternative to NB.

From our experience with the dataset we note that it seems that
class M has a similar, but narrower, distribution than class T.
This means that NBP predicts some points that are from class M to class T instead.
As we have discussed, the ability to correctly identify T in this case is a strength of the NB algorithm that we are not interested in preserving.

\subsection{Decision making}

\begin{table}
\centering
    \begin{tabular}{r|r|r}
    	Class & Very possible & Not impossible    \\ \hline
    	A & 20 & 79    \\ \hline
    	D & 22 & 70    \\ \hline
    	M & 39 & 111  \\ \hline
    	T & 112 & 205  \\ \hline
    	All & 261 & 261    \\ 
    \end{tabular}
    \caption{Summary of predictions for the escaped 
    salmon.}
    \label{tab:summary1}
\end{table}

In this section we focus on how to make the results useful in decision making, how to deliver value.
There are 261 escaped salmon for which we want to predict the location.
Table \ref{tab:summary1} shows one possible summary of the NBP model.
Here, two different cutoffs have been chosen to produce W, representing ``Very possible'' (fits well), and ``Not impossible'' (fits poorly but not extremely bad).
Then the W has been summed up across the 261 predictions tasks.
This table provides an overview at a glance of which location is more likely to have let more fish escape.
But if we use this table to make decisions, we are likely to make bad decisions.

Our main methodology for providing value from this imperfect dataset is to let the experts intervene several steps before the final prediction (exemplified in Table \ref{tab:summary1}).
We let the experts 
study the model evidence for each sample, and add
any extra information, before we together draw a conclusion about that fish.
This process often involves further investigation and data collection on the fish in question. 
For example, we can look up where, when, by whom, in what context, the fish was caught, lab notes, visual pictures of the fish or scales, or other data.
This has been a powerful approach in the project where we collect and use these data,
and we think it applies well to high stakes decision making based on imperfect data in general.

\begin{table}
\centering
    \begin{tabular}{r|r|r|r|r}
    	Fish & A & D & M & T    \\ \hline
    	F1 & 1 & 3 & 5 & 6    \\ \hline
    	F2 & 2 & 2 & 4 & 5    \\ \hline
    	F3 & 1 & 1 & 1 & 5  \\ \hline
    	F4 & 1 & 1 & 1 & 2    \\ 
    \end{tabular}
    \caption{The first rows of the graded model predictions. A 1 shows that the class does not fit at all, and a 5 or 6 shows that the class fits well.}
    \label{tab:grades1}
\end{table}

We make the model predictions available as a table of scores from 1 to 6, where 1 represents a very bad fit to the class, and 5 and 6 represent good fits.
See Table \ref{tab:grades1} for the first few fish.
For fish F3 it is clear that it comes from location T, and no further investigation is needed.
For fish F1, it is not clear whether we should pick M or T as both classes fit very well.
In this case, we can delve deeper into the model prediction for any given fish by investigating where its p-value comes from.
For each fish the p-value is an aggregate of p-values of several scales and several element compositions per scale.
Subject matter experts and statistical experts can then meet and build an understanding of which covariates and which scales are reliable, and come up with an informed conclusion for each fish.
In this application, some elements may not generalise well. 
This means that even if the cross-validation is very good the predictions may be wrong, because the escaped salmon for some elements may have different amounts than what they should have if they were sampled from a location.

\section{Discussion}

The approach we present extends to any variant of Naive Bayes where p-values can be computed or approximated.
For continuous covariates we can use other families of distributions than the Gaussian, for discrete covariates we can use any discrete distribution.
We can also introduce dependencies between the distributions, and we have not tried
non-parametric estimates.

We aim to extend this approach to functional data.
In addition to the element samples we have discussed here, we also have linescans.
Linescans are scans across the entire scales, giving a function $f(t)$ for each element. 
These functions are just higher dimensional covariates, and, after we figure out how to handle the data, it will be natural to use the NBP approach also here.

We think the approach could be extended to generative neural networks.
These models often have an explicit or implicit density of the generated observation.
However, there may be technical difficulties in
defining and approximating a reasonable p-value in extremely high dimensions.

\section{Supplementary materials}
See code examples at
\url{https://haakonbakkagit.github.io/btopic136.html} and \\
\url{https://haakonbakkagit.github.io/btopic137.html}, using
the R-package found at \\
\url{https://github.com/haakonbakkagit/NBPclassify}.
The R-package contains the dataset, and the code examples show how to load and analyse it.

\section{Acknowledgements}

Thanks to the Norwegian Veterinary Institute and Sporbarhet AS for funding and data collection.

\bibliographystyle{alpha}
%\bibliography{sample}

\section*{References}

This version is a very early version, and there are no references added in this version. Please feel free to email me if you know of especially relevant papers, or if you want to discuss the approach (haakon.c.bakka (at) gmail.com).

\end{document}